\documentclass{aastex}

\usepackage{graphicx}
\usepackage{dcolumn}
\usepackage{bm}

\newcommand{\dfrac}[2]{\displaystyle\frac{#1}{#2}}

\newcommand{\ME}{M$_{\mbox{E}}$}
\newcommand{\RE}{R$_{\mbox{E}}$}

\begin{document}


\title{Mass-Radius Relationships for Exoplanets II: \\
   Gr\"uneisen Equation of State for Ammonia}

\date{September 18, 2011; modified October 13, 2011 -- LLNL-JRNL-505357}

\author{Damian C. Swift}
\email{dswift@llnl.gov}
\affil{%
   Condensed Matter and Materials Division,
   Lawrence Livermore National Laboratory,
   7000 East Avenue, Livermore, California 94550, USA
}

\begin{abstract}
We describe a mechanical equation of state for NH$_3$,
based on shock wave measurements for liquid ammonia.
The shock measurements, for an initial temperature of 203\,K, 
extended to 1.54\,g/cm$^3$ and 38.6\,GPa.
The shock and particle speeds were fitted well with a straight line,
so extrapolations to higher compressions are numerically stable,
but the accuracy is undetermined outside the range of the data.
The isentrope through the same initial state was estimated,
along with its sensitivity to the Gr\"uneisen parameter.
Mass-radius relations were calculated for self-gravitating bodies of
pure ammonia, and for differentiated ammonia-rock bodies.
The relations were insensitive to variations in the Gr\"uneisen parameter,
indicating that they should be accurate for studies of planetary structure.
\end{abstract}


\maketitle

\section{Introduction}
Ammonia, NH$_3$, is a common molecule in ice giant planets, which appear to be
found widely throughout the galaxy \citep{Schneider2011}.
The equation of state (EOS) of ammonia is therefore important for
our understanding of planetary structures and their evolution,
potentially to pressures of order 1\,TPa for the base of the ice-rock
interface in icy exoplanets.
An accurate EOS for ammonia is also needed for studies of hypervelocity
impacts, such as meteoroid collisions with ice giants.
Furthermore, ammonia is a simple prototype for bonds occurring in chemical
explosives, for which densities from up to around twice that of zero-pressure
solids are of interest for shock initiation and detonation.

Although shock compression experiments have been performed on ammonia
to pressures of several tens of gigapascals \citep{Marsh1980},
the only equation of state readily available is SESAME 5520 \citep{Holian1984},
based on National Bureau of Standards gas phase data \citep{Johnson1982_5520},
and is tabulated to a maximum density of 0.765\,g/cm$^3$,
which is barely greater than the zero-pressure density for liquid ammonia.
Quasistatic compression experiments have been performed in which the density
and sound speed were measured along isotherms 
\citep{Abramson2008,Li2009}, but the highest pressures reported have reached only
a few gigapascals.

\section{Empirical Gr\"uneisen equation of state}
\begin{figure}
\begin{center}\includegraphics[width=\columnwidth]{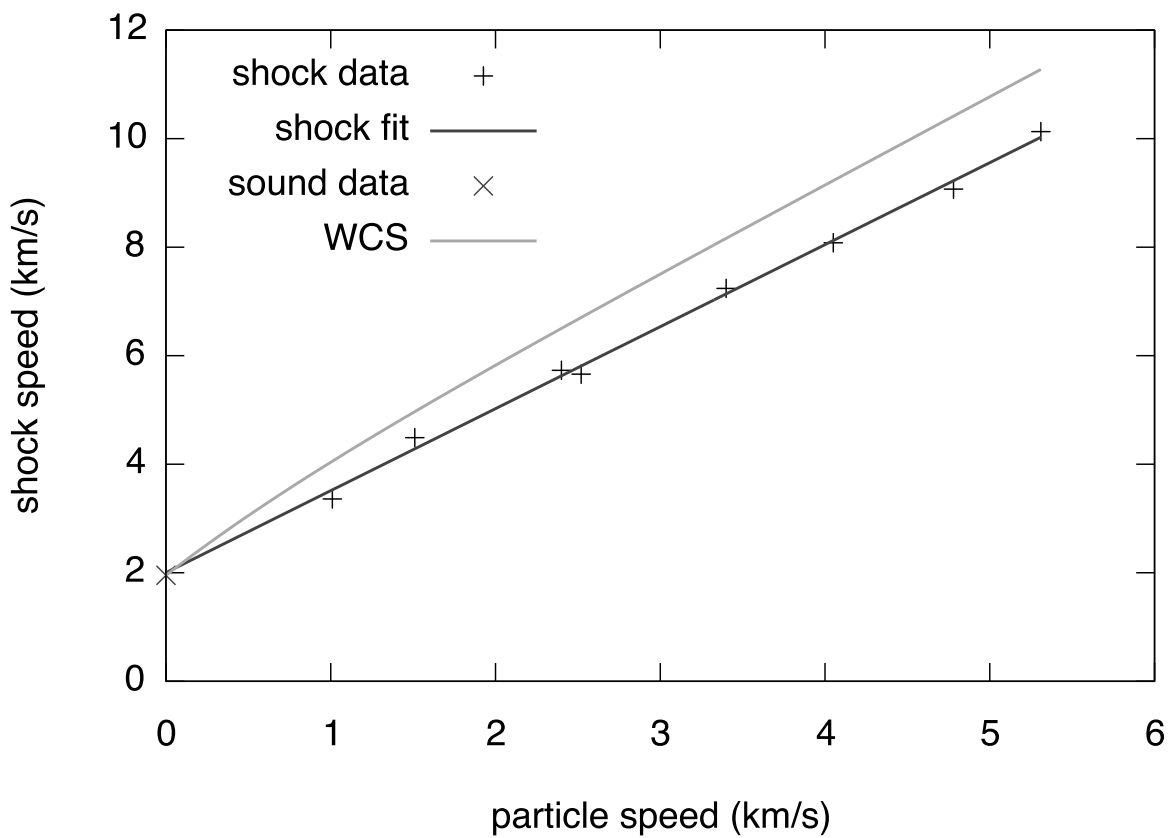}\end{center}
\caption{Principal shock Hugoniot of liquid ammonia (initial temperature 203\,K):
   experimental measurements and least-squares fit.
   The point at zero particle speed is the observed sound speed,
   which was not included in the fit.
   The curve labelled WCS is the Woolfolk-Cowperthwaite-Shaw
   universal liquid equation of state, whose sole fitting parameter is the
   sound speed at zero pressure.}
\label{fig:hugupus}
\end{figure}

Shock experiments have been reported previously
in which the shock and particle speeds $u_s$ and $u_p$ 
were measured for a range of shock
pressures, for liquid ammonia at an initial temperature of 203\,K
\citep{Marsh1980} (Fig.~\ref{fig:hugupus}).
The uncertainties in $u_s$ and $u_p$ were approximately 1\%.
These data can be fitted by a straight line fit
\begin{equation}
u_s = c_0 + s_1 u_p
\end{equation}
with
\begin{eqnarray*}
c_0 & = & 2.00 \pm 0.13 \quad (6.7\%) \\
s_1 & = & 1.511 \pm 0.039 \quad (2.6\%),
\end{eqnarray*}
where the standard errors shown are from the residual fitting error, neglecting
the uncertainty in measurement.
The experimental measurements exhibited a slightly curved trend,
but the number of points was not great enough to justify a higher-order fit.
Solving the Rankine-Hugoniot equations for a steady shock
\citep{Zeldovich1966} using the fitted parameters rather
than the individual shock measurements, the highest shock pressure
was 38.6\,GPa, giving a mass density of 1.54\,g/cm$^3$.
The observed sound speed in liquid ammonia at 203\,K is 1.9535\,km/s
\citep{Bowen1968}, which is consistent with the extrapolated Hugoniot data.
(Fig.~\ref{fig:hugupus})

Various universal EOS have been proposed for different classes of material.
It is interesting to compare with the `universal liquid EOS'
of \cite{Woolfolk1973}, whose only material-specific
parameter is the sound speed at zero pressure.
This EOS does not reproduce the shock data for ammonia, which is
softer and more linear than the universal EOS (Fig.~\ref{fig:hugupus}).

The fit to the shock Hugoniot can be used to predict the 
mechanical equation of state, using the Hugoniot as a reference curve 
\citep{McQueen1970}
\begin{eqnarray}
p(\rho,e) & = & p_r(\rho) + \Gamma(\rho)\left[e - e_r(\rho)\right] \\
p_r(\rho) & = & \dfrac{c_0^2\rho_r\rho(\rho-\rho_r)}{\left[\rho+s_1(\rho-\rho_r)\right]^2} \\
e_r(\rho) & = & e_0 + \frac 12 p_r(\rho)\left(\frac 1{\rho_r}-\frac 1\rho\right),
\end{eqnarray}
where $\rho_r$ is the initial density on the reference curve,
and $p_r(\rho)$ was derived for zero initial pressure, as here.
Other experiments are required to determine $\Gamma(\rho)$,
such as sound speed measurements on the Hugoniot, a shock Hugoniot from
a different initial state, or ramp compression.
However, $\Gamma$ can be estimated from the slope of the shock Hugoniot
as $2 s_1-1$, which is accurate for cubic crystals \citep{Skidmore1965}.
Thus $\rho_r = 0.725$\,g/cm$^3$ and $\Gamma\simeq 2.022$.

\begin{figure}
\begin{center}\includegraphics[width=\columnwidth]{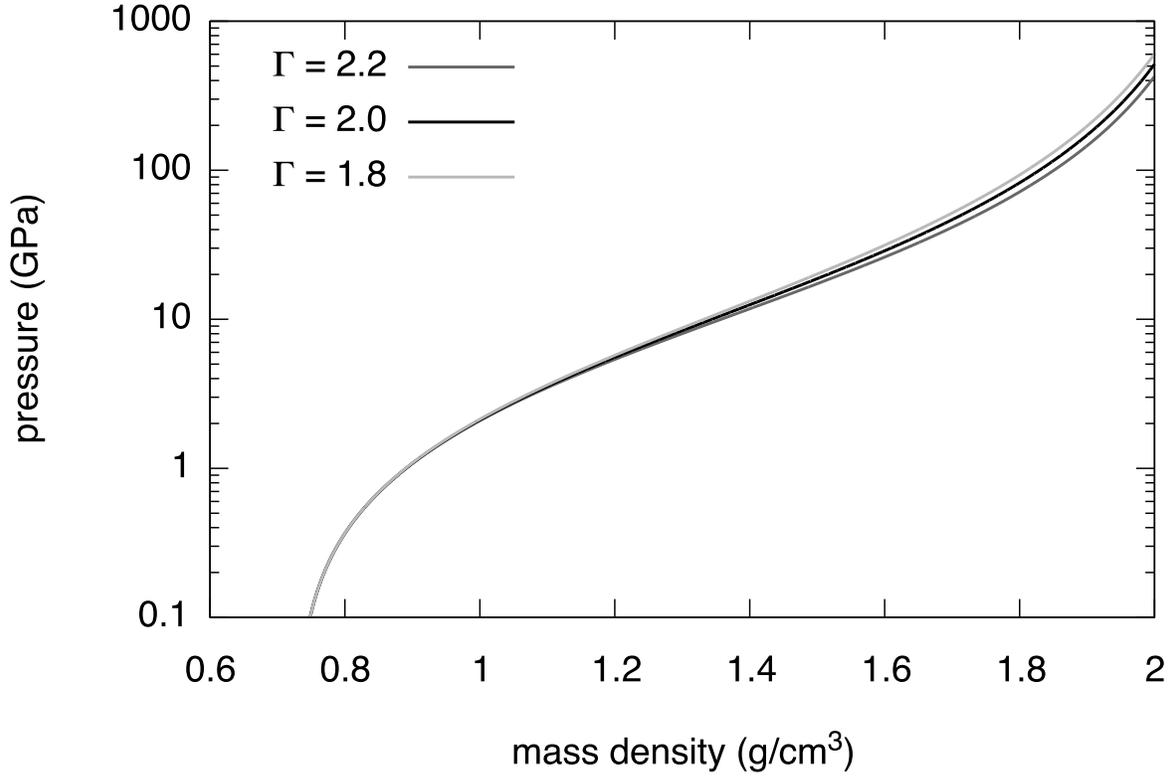}\end{center}
\caption{Principal isentrope of liquid ammonia (initial temperature 203\,K)
   deduced from mechanical equation of state fitted to principal shock
   Hugoniot, showing sensitivity to assumed Gr\"unseisen parameter.}
\label{fig:isendp}
\end{figure}

Given the mechanical EOS, the isentrope through any state can be calculated
by integrating the $-p\,dv$ work numerically \citep{Swift_genscalar_2008}.
Isentropes calculated from Gr\"unseisen EOS fitted to shock data typically
behave unphysically at high compression, where the assumption that the
Gr\"uneisen parameter is a function of density only breaks down.
For the ammonia fit, the breakdown occurred at a mass density of 2.145\,g/cm$^3$.
The isentrope was well-behaved to several terapascals, 
though its accuracy was undetermined.
The isentrope should be reasonably accurate at least up to the peak
compression in the shock data, which equates to 22.1\,GPa on the isentrope.
To investigate the sensitivity to the Gr\"uneisen parameter,
isentropes were calculated for the nominal value above,
and for values 10\%\ lower and higher.
With this variation in $\Gamma$, the pressure varied by 10\%\ at 20\,GPa,
rising to 25\%\ at 500\,GPa.
(Fig.~\ref{fig:isendp})

\section{Mass-radius relationships}
Mass-radius relationships were calculated using the deduced EOS,
for a self-gravitating body comprising pure ammonia and also for 
differentiated bodies consisting of a rocky core and an ammonia mantle,
using the numerical methods described previously \citep{Swift_mrexo_2011}.
Separate mass-radius curves were constructed for the nominal and
perturbed values of the Gr\"uneisen parameter.
In all cases, the temperature at the surface was taken to be 203\,K,
to match the initial state in the shock experiments.
The rocky core was modeled using an EOS for basalt, 
SESAME 7530 \citep{BarnesLyon1988_basalt},
as was done previously \citep{Swift_mrexo_2011}.

\begin{figure}
\begin{center}\includegraphics[width=\columnwidth]{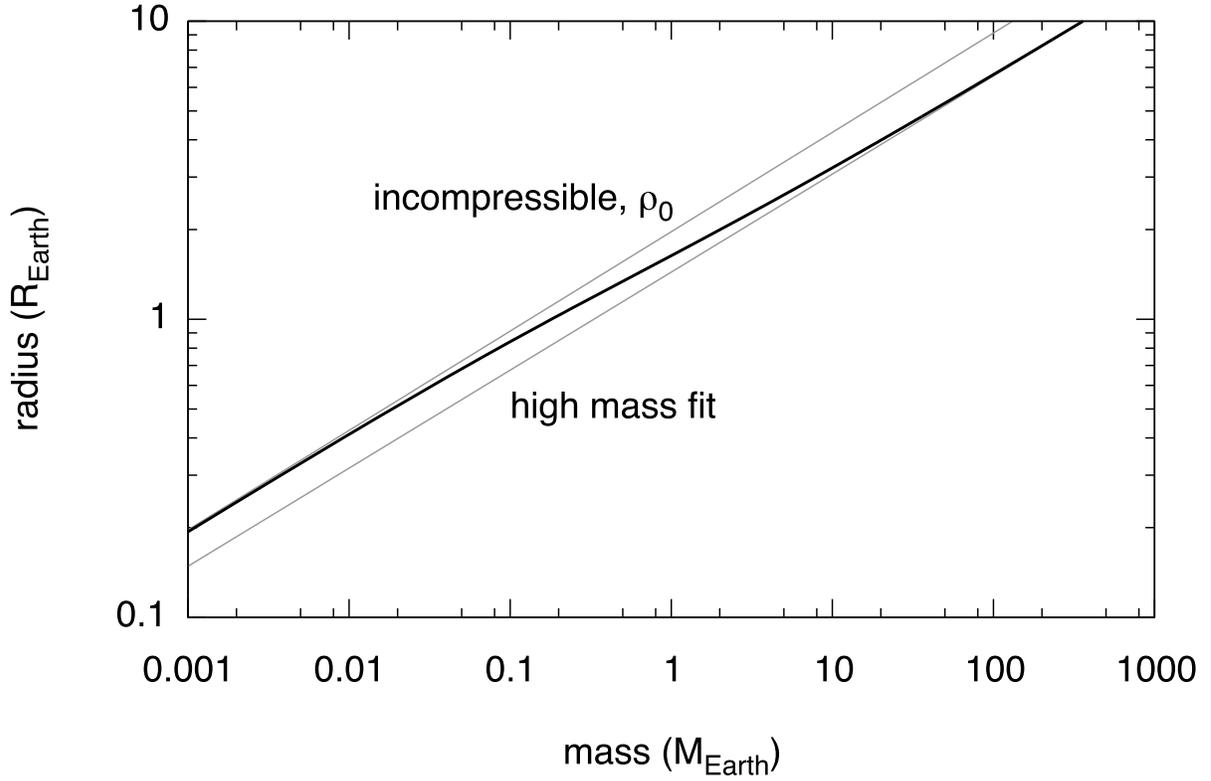}\end{center}
\caption{Mass-radius relation deduced from equation of state for
   liquid ammonia (surface temperature 203\,K),
   also showing relation for incompressible material and least-squares fit
   to the relation, which is dominated by high pressure behavior.}
\label{fig:mr}
\end{figure}

\begin{figure}
\begin{center}\includegraphics[width=\columnwidth]{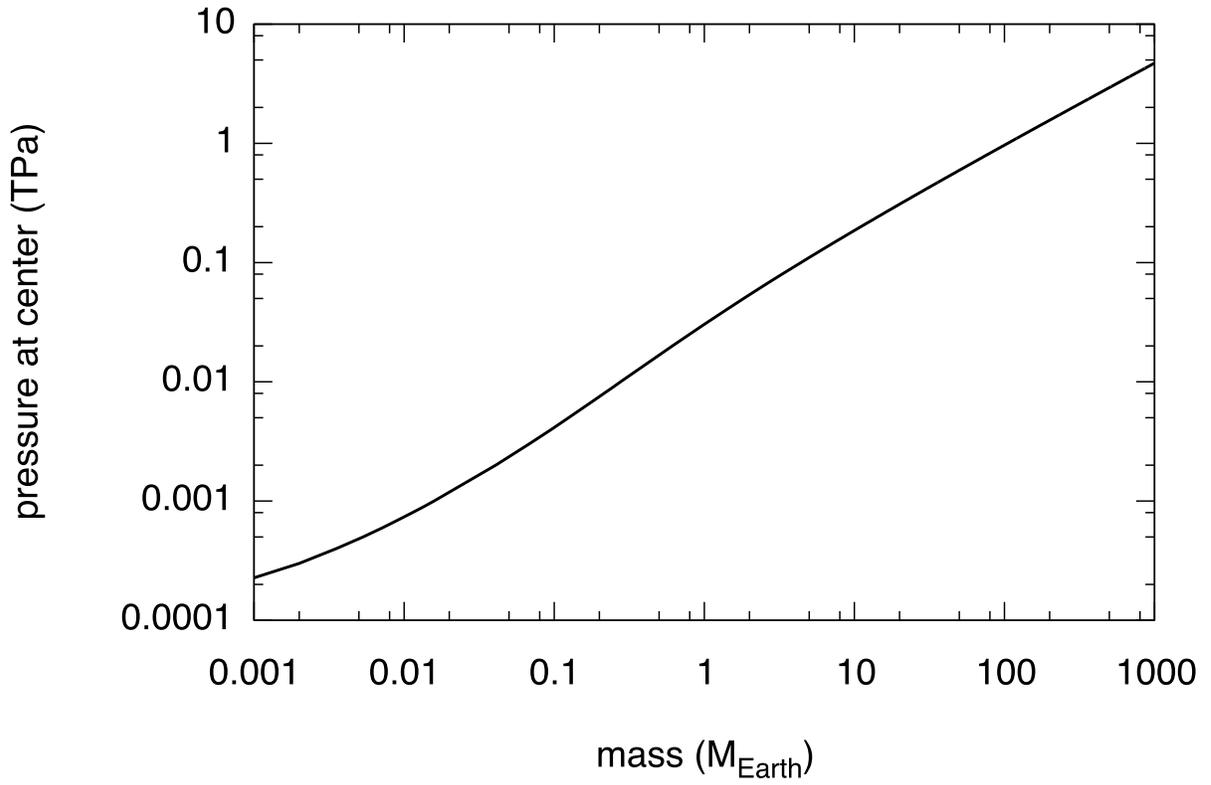}\end{center}
\caption{Variation of central pressure with mass, for liquid ammonia
   (surface temperature 203\,K).}
\label{fig:mp}
\end{figure}

\begin{figure}
\begin{center}\includegraphics[width=\columnwidth]{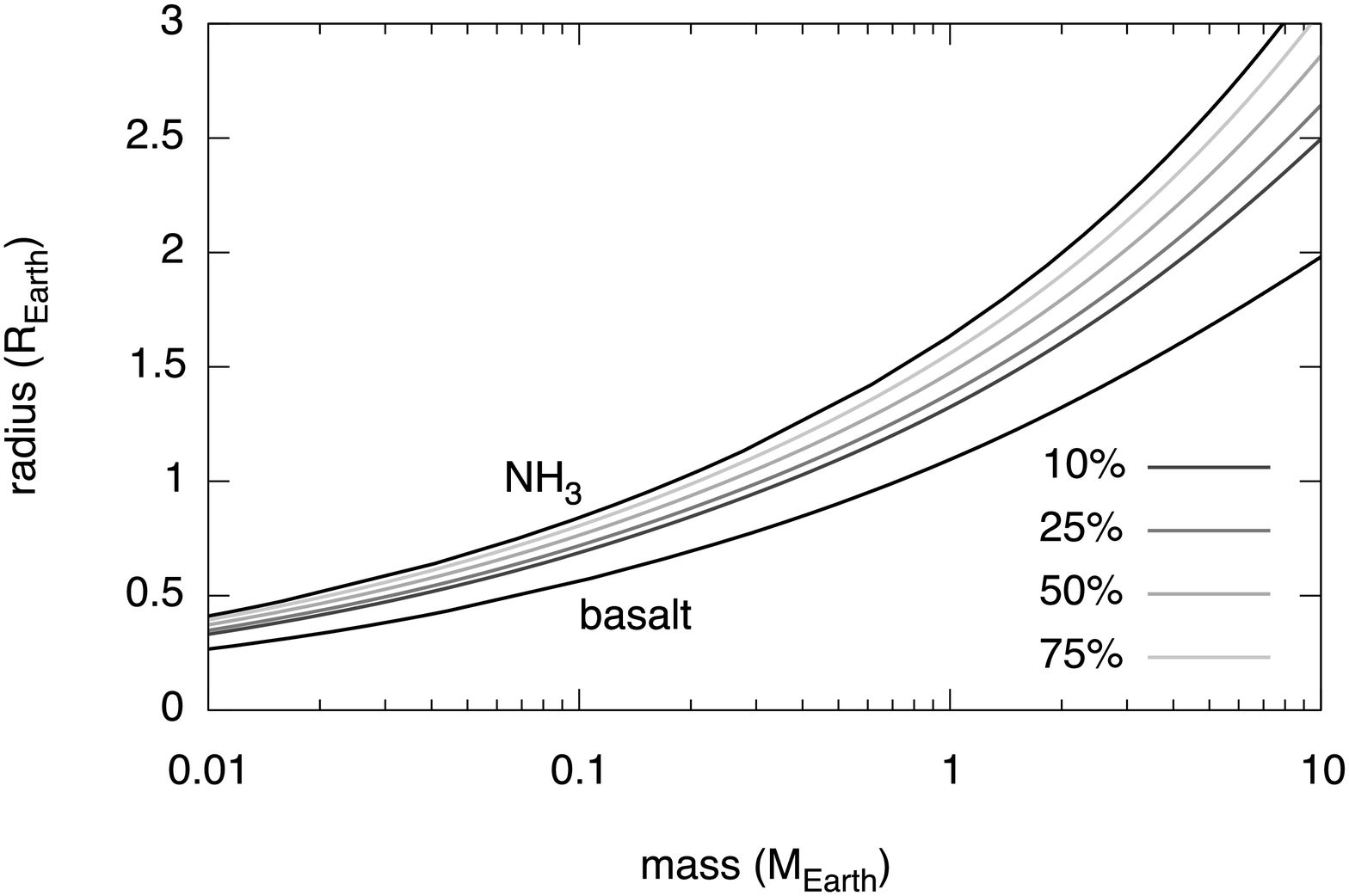}\end{center}
\caption{Mass-radius relations for differentiated ammonia-rock bodies.
   The percentages are the mass fraction of ammonia in the body.}
\label{fig:diffmr}
\end{figure}

The variations in $\Gamma$ made a negligible difference to the mass-radius
relations.
At high masses, the mass-radius relation for pure ammonia
asymptoted to a power-law behavior $R=\alpha M^\beta$ with
$\alpha = 1.4395 \pm 0.0005$ and $\beta  = 0.32889 \pm 0.00004$.
For an incompressible material,
\begin{equation}
M=\frac 43\pi r^3\rho_0,
\end{equation}
giving $\beta=1/3$.
The difference in the fitted value is small but significant;
$\alpha$ is considerably less than the incompressible value of
$\left(\frac 43\pi\rho_0\right)^{-1/3}$.
The mass-radius relation was also deduced using the SESAME EOS,
which matched that from the Gr\"uneisen EOS up to 0.1\,\ME,
above which point the extrapolation beyond the bounds of the table
gave unphysical behavior.
(Figs~\ref{fig:mr} to \ref{fig:diffmr}.)

The planetary radius for pure ammonia did not exhibit a maximum within the
range of masses investigated.
The range of compressions explored by the shock experiments was equivalent
to the central pressure in bodies of pure ammonia up to around 2/3\,\ME (1.5\,\RE).
However, the mass-radius relation is accurate for significantly larger
bodies, because the mass and volume are dominated by matter at much lower
pressures until the average density exceeds 1.5\,g/cm$^3$ or so:
approximately 4\,\ME, 2.5\,\RE, and a core pressure of 100\,GPa.
The relation may be accurate for even larger bodies, but it has not been
validated by EOS experiments.

\section{Conclusions}
The relation between shock and particle speeds in liquid ammonia appears
linear to within the scatter in the data up to pressures of at least 39\,GPa.
A Gr\"uneisen mechanical equation of state was constructed using 
the principal Hugoniot of initial state zero pressure and 203\,K as a reference,
and estimating the Gr\"uneisen parameter $\Gamma$ from the slope of the
Hugoniot.
Isentropes were calculated through the same state, the sensitivity to
$\Gamma$ rising with pressure.

Mass-radius relations were calculated for self-gravitating bodies consisting
of ammonia, and differentiated ammonia-rock mixtures.
The mass-radius relations were insensitive to variations in $\Gamma$,
indicating that the relations should be reliable for comparison to
planetary measurements, to central pressures substantially above those
reached in the shock experiments.

\section*{Acknowledgements}
This work was performed under the auspices
of the U.S. Department of Energy under contract DE-AC52-07NA27344.

\end{document}